\documentclass[twocolumn,showpacs,preprintnumbers,amsmath,amssymb]{revtex4}
\usepackage{epsfig,graphicx,dcolumn,bm,times,verbatim}

\begin{document}
\title{Response of degree-correlated scale-free networks to stimuli}
\author{Sheng-Jun Wang,$^1$ An-Cai Wu,$^1$ Zhi-Xi Wu,$^1$ Xin-Jian Xu,$^2$ and Ying-Hai Wang$^1$\footnote
{Electronic address: yhwang@lzu.edu.cn}}
\affiliation{$^1$Institute of Theoretical Physics, Lanzhou
University, Lanzhou Gansu 730000, China\\
$^{2}$Departamento de F\'{i}sica da Universidade de Aveiro,
3810-193 Aveiro, Portugal}

\date{\today}

\begin{abstract}
The response of degree-correlated scale-free attractor networks to
stimuli is studied. We show that degree-correlated scale-free
networks are robust to random stimuli as well as the uncorrelated
scale-free networks, while assortative (disassortative) scale-free
networks are more (less) sensitive to directed stimuli than
uncorrelated networks. We find that the degree-correlation of
scale-free networks makes the dynamics of attractor systems
different from uncorrelated ones. The dynamics of correlated
scale-free attractor networks result in the effects of degree
correlation on the response to stimuli.
\end{abstract}

\pacs{89.75.Hc, 87.18.Sn, 05.50.+q, 05.40.-a}

\maketitle

Many complex systems have the ability to react to low levels of
special stimuli, whereas, they can maintain their state when
exposed to high levels of other irrelevant stimuli \cite{Bar-Yam}.
If we take the units of response as nodes and the interactions
between responding units as edges, the structure of some these
systems can be described as complex networks. In neural networks
or social networks, for example, the nodes are individual neurons
or persons. It is an interesting problem that how one system have
both the sensitivity to the right stimuli and robustness in the
face of the wrong one. And the problem is also important for
designing large artificial complex systems. The source of the
ability of networked complex systems to incorporate the two
complementary attributes have been investigated using network
models. It was shown that the power-law shape degree distributions
of networks give rise to the sensitivity and robustness in a
system \cite{Bar-Yam}.

The topology of real networks is also characterized by degree
correlation \cite{Vazquez, Pastor-Satorras, Newman-l}. In a
network with degree correlation, there exist certain relationships
between network nodes. The degree correlations are often named
respectively as ``assortative mixing'', i.e. a preference for
high-degree nodes to attach to other high-degree nodes, while ``
disassortative mixing'' --- high-degree nodes attach to low-degree
ones \cite{Newman-l}. It has been pointed out that the existence
of degree correlations among nodes is an important property of
real networks \cite{Xulvi-Brunet,
Newman-E,Vaquez-2,Capocci,Newman-3,Berg,Goh,Maslov,Krapivsky,Dorogovtsev}.
The percolation \cite{Newman-l} and disease spreading
\cite{Boguna} on correlated networks have been studied. And more
effects of degree correlation on network structure and function
have attracted attention \cite{Gallos,Bianconi,Fronczak}.
Therefore, the extension of previous results for uncorrelated
network model about responding to stimuli is necessary.

In this paper, we study the response of degree-correlated
scale-free networks to stimuli following the work contributed by
Bar-Yam and Epstein \cite{Bar-Yam}. Numerical investigation
reveals that the dynamical process of the evolution of attractor
systems on correlated scale-free networks is different from
uncorrelated networked systems. The special dynamics of correlated
attractor systems result in the different responding behavior from
uncorrelated systems. The degree-correlated scale-free network is
robust in the face of wrong stimuli as uncorrelated networks. In
assortative networks, the sensitivity to right stimuli is
enhanced. While in the disassortative networks the sensitivity to
right stimuli is weaker than uncorrelated networks. And, the
relation between the sensitivity to stimuli and the degree of
correlation is not monotonic.

We consider the method for modelling the response of complex
systems proposed in \cite{Bar-Yam}. We use a model of attractor
networks \cite{Hopfiled, Amit}, where the node states $s_i=\pm 1,
i\in \{1,\cdots, N \}$ are binary. The state of the system is the
set of node states $\{ s_i \}$. The dynamical equations of the
attractor system are
\begin{equation}\label{eq:evolve}
s_i(t+1)=\textrm{sign}(\sum_{j=1}^{N} J_{ij}s_j(t)),
\end{equation}
with symmetric influence matrix $J_{ij}$. Using the Hebbian
imprinting rule
\begin{equation}\label{eq:influence}
J_{ij}=\sum_{\alpha} c_{ij} s_i^{\alpha}s_j^{\alpha},
\end{equation}
we can set the states $\{ s_i^{\alpha}\}_{\alpha=1,\cdots,n}$ as
the stable states of the network dynamics (attractor). $c_{ij}$ is
the entry of the symmetric adjacent matrix which is equal to 1
when node $i$ connects to node $j$, and zero otherwise. An
attractor is stable to perturbation and thus can represent a
functional state of systems. In simulations, we randomly choose
two functional states of the system $\{ s_i^{\alpha} \}_{\alpha=1,
2}$, and the influence is $J_{ij}=\sum_{\alpha=1}^2 c_{ij}
s_i^{\alpha}s_j^{\alpha}$. External stimuli are modelled by
changing the signs of a specified set of nodes. When the states of
some nodes are flipped, the system either evolves back to its
initial state or switches to other stable system states. The
response of networked systems is described as a process of
switching between attractors. The size of the basin of attraction,
the number of nodes whose states can be changed before the
dynamics of the network fails to bring the system back to its
original state, indicates the degree of stability of the system.
We calculate the size of the basin of attraction in different
cases of stimuli to reveal the sensitivity and robustness of the
network model.

Generally, degree-correlated networks can be generated from
uncorrelated ones by means of reshuffling method proposed in
\cite{Xulvi-Brunet}. Starting from a given network, at each step
two edges of the network are chosen at random. The four nodes
attached to the two edges are ordered with respect to their
degrees. Then with probability $p$, the edges are rewired in such
a way that one edge connects the two nodes with the smaller
degrees and the other connects the two nodes with the larger
degrees; otherwise, the edges are randomly rewired. In the case
when one or both of these new edges already existed in the
network, the step is discarded and a pair of other edges is
selected. A repeated application of the rewiring step leads to an
assortative networks. For producing disassortative networks, we
modify the way for building new edges used in above reshuffling
method as that the node of the largest degree connects to the
nodes of the smallest degree and two other nodes are connected. It
is worth noting that the algorithm does not change the degree
distribution in the given network \cite{Xulvi-Brunet}.

Before investigating the effect of the degree correlation on the
response, we review the results on uncorrelated attractor networks
\cite{Bar-Yam}, where the system was characterized by the
scale-free networks which have the power-law shape degree
distribution $P(k)\sim k^{-\gamma}$. The size of the basin of
attraction for two kinds of stimuli, namely, the random stimuli
(randomly chosen nodes are flipped) and the directed stimuli
(means flipping sequentially the nodes of greatest degree) were
studied on scale-free attractor network systems. The relation
between the size of the basin of attraction for random stimuli
$b_r$ and directed stimuli $b_m$, which are all normalized by
network size $N$, are derived:
\begin{equation}\label{eq:uncorrelated}
b_m=b_r^{(\gamma-1)/(\gamma-2)}.
\end{equation}
The derivation was based on a assumption that the response of
attractor networks occurs if the sum of edges coming from
stimulated nodes exceeds a threshold which is the same for both
random and directed stimuli. For Barab\'asi-Albert (BA) scale-free
networks \cite{BA}, the distribution exponent $\gamma=3$ and thus
$b_m=b_r^2$. So the scale-free networks are robust to random
stimuli and sensitive to directed stimuli.

\begin{figure}
\centerline{\epsfxsize=9cm \epsffile{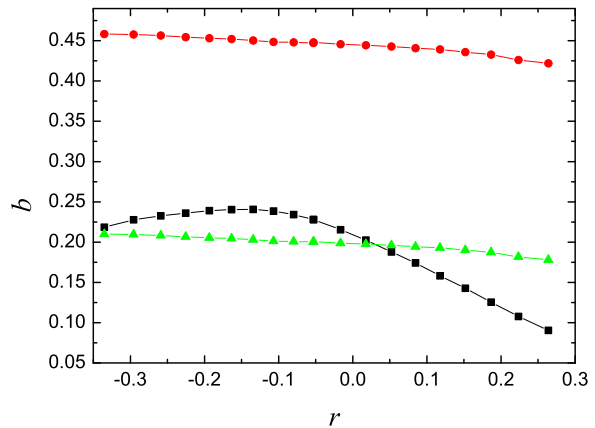}}
\caption{\label{fig:basin}(Color online) The size of attractor
basin of scale-free networks as a function of Pearson correlation
coefficient $r$ in the case of directed (square) and random
stimuli (circle). All networks have the same network size $N=1000$
and average degree $\langle k \rangle=20$. Each curve is an
average of 1000 realizations. The predicted curve of $b_m'$
calculated using Eq.(\ref{eq:uncorrelated}) is shown as the curve
with triangles.}
\end{figure}

Let us first calculate the average size of the basin of attraction
for random stimuli $b_r$ and directed stimuli $b_m$ on
degree-correlated BA networks. According to \cite{Bar-Yam}, we use
the network size $N=1000$ and average degree $\langle k
\rangle=20$ in all simulations. Figure \ref{fig:basin} shows the
average size of attractor basin versus the degree of correlation
which is quantified by the Pearson correlation coefficient $r$
\cite{Newman-l}. To compare with uncorrelated case, in Fig.
\ref{fig:basin} we also plot the predicted size of the attractor
basin for directed stimuli $b_m'$ which is calculated using the
size of the attractor basin for random stimuli $b_r$ following Eq.
(\ref{eq:uncorrelated}). Restricted by the reshuffling method, we
can not generate networks with strong degree correlation
$|r|\rightarrow 1$ \cite{Xulvi-Brunet}. In simulations, the region
of the Pearson correlation coefficient $r$ is about from $-0.3$ to
$0.3$. Although the region is small, it nearly covers all the
values of the Pearson correlation coefficient $r$ of realistic
complex networks shown in \cite{Newman-l}. Therefore, we interest
in systems with the Pearson correlation coefficient belonging to
the region about from -0.3 to 0.3.

In Fig. \ref{fig:basin} we can see the effects of the degree
correlation of scale-free networks on the size of the basion of
attraction. Comparing the size of attractor basin $b_m^{'}$
predicted using Eq.(\ref{eq:uncorrelated}) (the curve with
triangles) with the size obtained by computer simulations (the
curve with squares), one can see that the relation between the
size of attractor basin for random stimuli $b_r$ and directed
stimuli $b_m$ derived in uncorrelated case is not satisfied in
correlated scale-free networks. When $r \approx 0$ the numerical
result of the attractor basin for directed stimuli $b_m$ is
identical with the prediction of uncorrelated networks
$b_m'$.\cite{r0} For assortative case $r>0$, the basin of
attraction for directed stimuli is less than the value of
uncorrelated network. This means that the assortative scale-free
network is more sensitive to directed stimuli than uncorrelated
scale-free networks. For disassortative case, the size of
attractor basin undergoes a non-monotonic process with the
variance of Pearson correlation coefficient. The sensitivity of
disassortative scale-free networks is weaker than uncorrelated
systems. The size of the basin of attraction for random stimuli
$b_r$ decreases monotonically with the increase of $r$. And the
slope is small. The robustness of scale-free networks to random
stimuli retains when these networks are degree correlated.

To understand the underlying mechanism of the effect of degree
correlation on response, we analyze the dynamics of attractor
networks. We assume that there are $n$ functional states in an
attractor system. Substitute of Eq. (\ref{eq:influence}) into Eq.
(\ref{eq:evolve}) gives
\begin{align}
s_i(t+1)&=\textrm{sign}(\sum_{j=1}^{N}\sum_{\alpha=1}^{n}c_{ij}s_{i}^{\alpha}s_{j}^{\alpha}s_j(t)) \nonumber\\
&=\textrm{sign}(\sum_{\alpha=1}^{n}s_i^{\alpha}\sum_{j \in
G_i}s_j^{\alpha}s_j(t)),
\end{align}
where $G_i$ is the set of nodes adjacent to node $i$ (the
neighbors of node $i$). We use the functional state $\{s_i^1\}$ as
the original system state, and the stimulated system state is
denoted as $\{s_i^{\beta}\}$. Thus the first step of the evolution
is like
\begin{equation}\label{eq:evolve-first}
s_i(1)=\textrm{sign}(s_i^1\sum_{j \in
G_i}s_j^1s_j^{\beta}+\sum_{\alpha=2}^ns_i^{\alpha}\sum_{j\in
G_i}s_j^{\alpha}s_j^{\beta}).
\end{equation}
The functional states $\{s_i^{\alpha}\}_{\alpha=2,\cdots,n}$ are
uncorrelated with the stimulated state $\{ s_i^{\beta}\}$, since
the functional states are chosen at random. Thus the second term
in the bracket at the right side of Eq. (\ref{eq:evolve-first}) is
approximately equal to 0, and this term can be taken as noise
\cite{Hopfiled}. For an arbitrary node $i$, if much less than half
nodes in $G_i$ are flipped by the stimulus, then $s_i(1)=s_i^1$;
if much more than half nodes in $G_i$ are flipped,
$s_i(1)=-s_i^1$. In general, the fraction of flipped nodes in
$G_i$ increases as stimuli are enhanced. Because of the influence
of noise, when the fraction of flipped nodes in $G_i$ is near but
less than $0.5$, the node $i$ choose a state $s_i^1$ or $-s_i^1$
at random.

In the case of uncorrelated networks, for both random and directed
stimuli, the fraction of flipped nodes in neighbors of each node
is equal to the fraction $f$ of edges coming from flipped nodes in
a network. This property determines a critical condition for
uncorrelated systems responding to stimuli: near half edges in a
network come from the stimulated nodes. We obtained the critical
value of $f$ on the system with two functional states by numerical
simulation, which is $f_c=0.46$ for both random and directed
stimuli. When stimuli are large enough to satisfy the critical
condition, all nodes in uncorrelated networks choose their states
at random with the help of noise term. Then, the system state $\{
s_i(1) \}$ becomes a random state, and evolves to one of
attractors randomly. The analysis of the above property gives an
insight of the dynamics of uncorrelated networks that the
uncorrelated networks responds to both kinds of stimuli as a
whole.

\begin{figure}
\centerline{\epsfxsize=9cm \epsffile{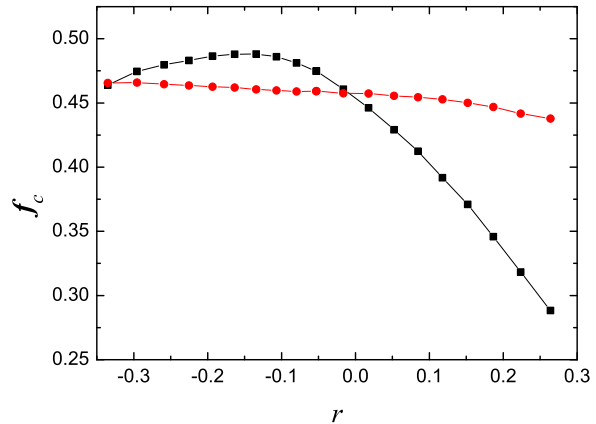}} \caption{
\label{fig:fraction} (Color online) The critical value of the
number of edges attached to flipped nodes as a function of Pearson
correlation coefficient in the case of directed (square) and
random (circle) stimuli. Each curve is an average of 1000
realizations.}
\end{figure}

Figure \ref{fig:fraction} shows numerical results of the critical
fraction of edges attached to stimulated nodes versus the Pearson
correlation coefficient of reshuffling scale-free networks. When
networks are degree-correlated, the difference between the
critical fraction $f_c$ for random stimuli and directed stimuli is
remarkable. The result shows that the mentioned assumption used
for deriving Eq.(\ref{eq:uncorrelated}) in \cite{Bar-Yam} is not
appropriate for degree-correlated scale-free networks. In Fig.
\ref{fig:fraction}, one can note that the critical fraction $f_c$
for random stimuli varies slightly. Under random stimuli, for
correlated scale-free networks, the fraction of flipped nodes in
the neighbor of each node is approximately equal to the fraction
$f$ of edges coming from flipped nodes in a network. The dynamics
of degree-correlated scale-free networks under random stimuli have
the same characteristic as uncorrelated networks: the attractor
systems respond to random stimuli as a whole. Under directed
stimuli, the variation of $f_c$ versus the Pearson correlation
coefficient indicates that the dynamics of directed stimulated
attractor networks are affected seriously by degree-correlation.

\begin{figure}
\centerline{\epsfxsize=9cm \epsffile{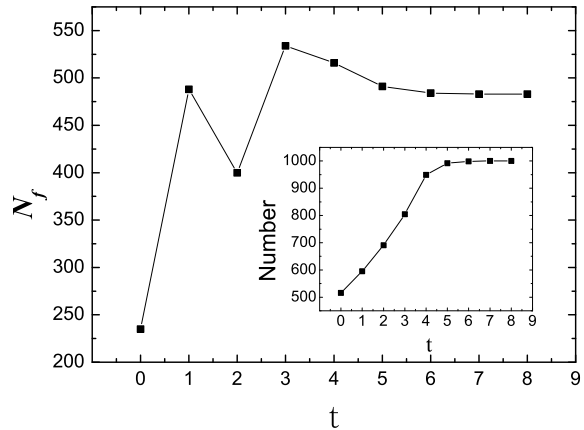}}
\caption{\label{fig:dyn-uncor} The number of flipped nodes in the
process of the evolution of the uncorrelated system. Inset: the
number of nodes whose state $s_i(t)$ is the same as $s_i^2$.}
\end{figure}

Next we numerically investigate the dynamical process of the
evolution of the attractor system in the case of directed stimuli,
and reveal the underlying mechanism of the effect of degree
correlation. To do this, we give a directed stimuli with size
equal to 235 to a realization of the uncorrelated network. The
stimulus is larger than the average attractor basin for
uncorrelated scale-free attractor systems given in Fig.
\ref{fig:basin} which is equal to $215(\pm 12)$. In Fig.
\ref{fig:dyn-uncor} the dynamical process of the evolution of the
system is represented by the number of flipped nodes ($N_f$). At
the first step of the evolution, the number of the flipped nodes
is 488, which is near half of the network size. And then the
system evolves to another imprinted functional state, as shown in
the inset of Fig. \ref{fig:dyn-uncor}. The evolution shows that
the uncorrelated scale-free networks response to directed stimuli
as a whole, as the above analysis.

\begin{figure}
\centerline{\epsfxsize=9cm \epsffile{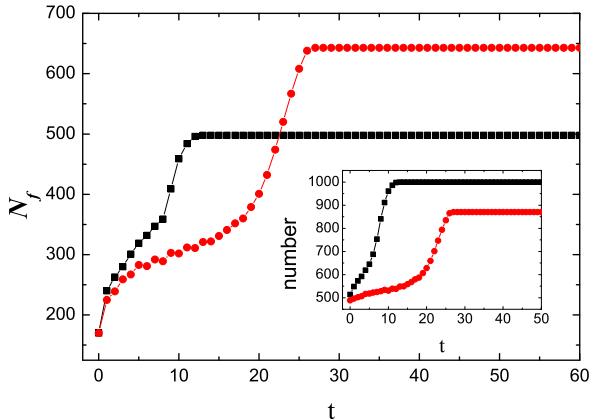}}
\caption{\label{fig:dyn-assor} (Color online) The number of
flipped nodes in the process of the evolution of two assortative
systems with $r=0.13$ (square) and $r=0.15$ (circle). Inset: the
number of nodes whose state $s_i(t)$ is the same as $s_i^2$.}
\end{figure}

For assortative networks, we give a directed stimulus with the
size 170 to attractor systems. Although the size of the stimuli is
smaller than the mentioned average attractor basin of uncorrelated
networks, the system responds to the stimulus with the process of
the change of the system state, as shown in Fig.
\ref{fig:dyn-assor}. We note that the number of flipped nodes
increases gradually. In contrast with uncorrelated scale-free
networks, the evolution shows that the assortative scale-free
network system does not make response as a whole. In assortative
scale-free networks, a group of nodes of large degree
preferentially connect to the nodes of greatest degrees, i.e.
stimulated nodes, and thus they are easier to get the condition
for changing their states. So the set of flipped nodes can be
extended by the assortative mixing. The assortative scale-free
network system evolves as a hierarchical cascade \cite{Barthelemy}
that progresses from higher to lower degree classes. Therefore the
basin of attraction of assortative network system decreases and
the system is more sensitive to directed stimuli.

With the increase of Pearson correlation coefficient, the cluster
coefficient of assortative networks are increased by the degree
based reshuffling steps \cite{Xulvi-Brunet}. The cluster property
also effects the dynamics of assortative scale-free networks. In
Fig. \ref{fig:dyn-assor} we show two numerical simulations with
different types of dynamics. For one kind of dynamics (square),
the stable system states are the functional states imprinted by
Hebbian rule, as the uncorrelated networks. The upper curve
(square) in the inset of Fig. \ref{fig:dyn-assor} shows that a
system evolves into the second functional state. For another kind
of dynamics (circle), the stable system state at the end of
evolution is not the imprinted functional state. The lower curve
(circle) in the inset of Fig. \ref{fig:dyn-assor} shows the
discrepancy. In this kind of systems, cluster forms between
stimulated nodes which have a high density of edges within them,
with a lower density of edges between other groups of nodes. So
these stimulated nodes hold their states on $-s_{i}^{1}$.
Additionally, the state of some low-degree nodes which connect
tightly to the cluster is also held. These nodes held by the
cluster structure result in the difference between the system
state and the imprinted functional state. There is a critical
value $r_c$, for the networks used in simulations $r_c=0.32$,
below which two types of dynamics are possible (and larger the
value of $r$ is, more frequently the second type of dynamics
occur), while above which systems only responde to stimuli by the
second type of dynamics. Because of the cluster property of
assortative networks, too larger assortative mixing is not
expected for response of networks. In the limit of $r \rightarrow
1$, networks disintegrate into isolated clusters, each of them
consists of nodes with certain degree $k$. Directed stimuli cannot
induce these systems to change their functional states, but only
change few clusters and leave the other nodes on their initial
states.

\begin{figure}
\centerline{\epsfxsize=9cm \epsffile{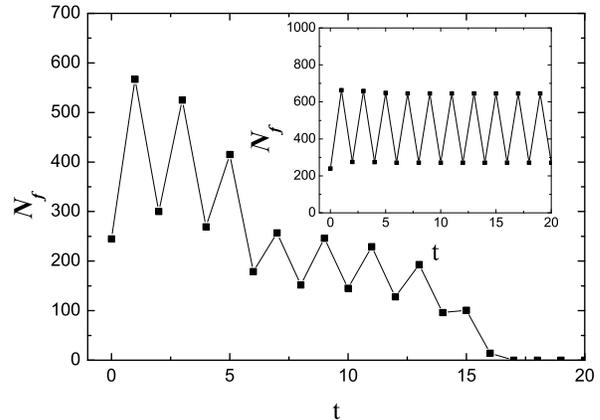}}
\caption{\label{fig:dyn-disas} The number of flipped nodes in the
process of disassortative system, $r=-0.16$. Inset: system with
$r=-0.30$. The size of stimuli is 245.}
\end{figure}

For the disassortative system, we choose a reshuffling scale-free
network realization with Pearson correlation coefficient $r=-0.16$
which has the lowest sensitivity to directed stimuli as shown in
Fig. \ref{fig:basin}. We give the disassortative network a
directed stimulus with size 245 which is larger than the average
attractor basin of the uncorrelated scale-free networks. Fig.
\ref{fig:dyn-disas} shows the dynamical process of the evolution
of the system. Although more than half of nodes flip their states
at the first step, the system state is attracted into the original
functional state. In disassortative networks, nodes with large
degrees preferentially connect to the nodes with small ones. Under
directed stimuli, the fraction of stimulated nodes in the
neighbors of the nodes in middle degree class is less than the
fraction of the edge coming from stimulated nodes. Thus, more
nodes need to be stimulated than uncorrelated systems for inducing
the system into random state, and the basin of attraction of
disassortative system extends.

For larger disassortative mixing systems, the second imprinted
functional state cannot be reached. The inset of Fig.
\ref{fig:dyn-disas} shows the dynamical process of evolution of a
network realization with $r=-0.30$. The system is induced into
stable oscillation state, which is established by the interaction
between large and small nodes. The system with large
disassortative mixing property is easier to responde the directed
stimuli by evolving into stable oscillation states. This
structural property leads to the non-monotonic behavior of
sensitivity versus Pearson correlation coefficient shown in Fig.
\ref{fig:basin}. Additionally, it is notable that the too large
disassortative degree correlation also destroys the ability of
systems to responde directed stimuli with imprinted functional
states, as the too large assortative degree correlation.

In summary, we have studied the effect of the degree correlation
on the response of scale-free networks to stimuli. Correlated
scale-free networks retain the robustness to random stimuli. In
the region of Pearson correlation coefficient in which we
interest, assortative scale-free networks are more sensitive to
directed stimuli than uncorrelated ones; and the sensitivity of
scale-free networks are weaken when networks are disassortative.
We found that the effects of degree correlation result from the
properties of the dynamics of degree-correlated network systems.
Uncorrelated networks responde to stimuli as a whole. While the
degree correlation of a network destroys the identical critical
condition of all nodes for the response to directed stimuli.
Assortative scale-free networks reduce the need on the size of
directed stimuli to be responded via a cascade that progresses
from higher to lower degree classes. The disassortative
correlation extends the size of the basin of attraction by the
nodes in middle degree class which has less stimulated neighbors
and stay on initial state. But the response of too large
assortative and disassortative scale-free networks is destroyed by
the structure property, and imprinted functional states cannot be
reached. Since many realistic complex networks have both
scale-free and degree-correlated properties, the intuitive
description of the dynamics might contribute to understanding of
the attributes of realistic networks.

This work was supported by the Fundamental Research Fund for
Physics and Mathematics of Lanzhou University under Grant No.
Lzu05008. X.-J. Xu acknowledges financial support from FCT
(Portugal), Grant No. SFRH/BPD/30425/2006.


\begin{references}

\bibitem{Bar-Yam}
Y. Bar-Yam and I. R. Epstein, Proc. Natl. Acad. Sci. USA
\textbf{101}, 4341 (2004).

\bibitem{Vazquez}
A. V\'{a}zquez and Y. Moreno, Phys. Rev. E \textbf{67}, 015101(R)
(2003).

\bibitem{Pastor-Satorras}
R. Pastor-Satorras, A. V\'{a}zquez and A. Vespignani, Phys. Rev.
Lett. \textbf{87}, 258701 (2001).

\bibitem{Newman-l}
M. E. J. Newman, Phys. Rev. Lett. \textbf{89}, 208701 (2002).

\bibitem{Xulvi-Brunet}
R. Xulvi-Brunet and I. M. Sokolov, Phys. Rev. E \textbf{70},
066102 (2004).

\bibitem{Newman-E}
M. E. J. Newman, Phys. Rev. E \textbf{67}, 026126 (2003).

\bibitem{Vaquez-2}
A. V\'{a}zquez, M. Bogu\~{n}\'{a}, Y. Moreno, R. Pastor-Satorras
and A. Vespignani, Phys. Rev. E \textbf{67}, 046111 (2003).

\bibitem{Capocci}
A. Capocci, G. Caldarelli, and P. De Los Rios, Phys. Rev. E
\textbf{68}, 047101 (2003).

\bibitem{Newman-3}
M. E. J. Newman and J. Park, Phys. Rev. E \textbf{68}, 036122
(2003).

\bibitem{Berg}
J. Berg and M. L\"{a}ssig, Phys. Rev. Lett. \textbf{89}, 228701
(2002).

\bibitem{Goh}
K.-I. Goh, E. Oh, B. Kahng, and D. Kim, Phys. Rev. E \textbf{67},
017101 (2003).

\bibitem{Maslov}
S. Maslov and K. Sneppen, Science \textbf{296}, 910 (2002).

\bibitem{Krapivsky}
P. L. Krapivsky and S. Redner, Phys. Rev. E \textbf{63}, 066123
(2001).

\bibitem{Dorogovtsev}
S. N. Dorogovtsev, Phys. Rev. E \textbf{69}, 027104 (2004).

\bibitem{Boguna}
M. Bogu\~{n}\'{a} and R. Pastor-Satorras, Phys. Rev. E
\textbf{66}, 047104 (2002).

\bibitem{Gallos}
L. K. Gallos and P. Argyrakis, Phys. Rev. E \textbf{72}, 017101
(2005).

\bibitem{Bianconi}
G. Bianconi and M. Marsili, Phys. Rev. E \textbf{73}, 066127
(2006).

\bibitem{Fronczak}
A. Fronczak and P. Fronczak, Phys. Rev. E \textbf{74}, 026121
(2006).

\bibitem{Hopfiled}
J. J. Hopfield, Proc. Natl. Acad. Sci. USA \textbf{79}, 2554
(1982).

\bibitem{Amit}
D. J. Amit, H. Gutfreund and H. Sompolinsky, Phys. Rev. Lett.
\textbf{55}, 1530 (1985).

\bibitem{BA}
A.-L. Barab\'asi and R. Albert, Science \textbf{286}, 509 (1999).

\bibitem{r0}
For $r\approx 0$, we have checked that the size of the attractor
basin does not change when using neutral shuffled networks instead
of BA networks.

\bibitem{Barthelemy}
M. Barth\'{e}lemy, A. Barrat, R. Pastor-Satorras and A.
Vespignani, Phy. Rev. Lett. \textbf{92}, 178701 (2004).

\end{references}
\end{document}